\documentclass[%
 reprint,
superscriptaddress,
nofootinbib,
 amsmath,amssymb,
 aps,
]{revtex4-1}
\usepackage{graphics,epsfig,psfrag}
\usepackage{color}
\usepackage{graphicx}
\usepackage{dcolumn}
\usepackage{bm}
\usepackage{mathrsfs}

\begin{document}

\preprint{APS/123-QED}


\title{Shadow of a black hole at cosmological distance}

\author{Gennady S. Bisnovatyi-Kogan}
\email{gkogan@iki.rssi.ru}
\affiliation{Space Research Institute of Russian Academy of Sciences, Profsoyuznaya 84/32, Moscow 117997, Russia}
\affiliation{National Research Nuclear University MEPhI (Moscow Engineering Physics Institute), Kashirskoe Shosse 31, Moscow 115409, Russia}
\affiliation{Moscow Institute of Physics and Technology, 9 Institutskiy per., Dolgoprudny, Moscow Region, 141701, Russia}%

\author{Oleg Yu. Tsupko}
\email{tsupko@iki.rssi.ru}
\affiliation{Space Research Institute of Russian Academy of Sciences, Profsoyuznaya 84/32, Moscow 117997, Russia}


\date{\today}



\begin{abstract}
Cosmic expansion is expected to influence on the size of black hole shadow observed by comoving observer. Except the simplest case of Schwarzschild black hole in de Sitter universe, analytical approach for calculation of shadow size in expanding universe is still not developed. In this paper we present approximate method based on using angular size redshift relation. This approach is appropriate for general case of any multicomponent universe (with matter, radiation and dark energy). In particular, we have shown that supermassive black holes at large cosmological distances in the universe with matter may give a shadow size approaching to the shadow size of the black hole in the center of our galaxy, and present sensitivity limits.


\pacs{?????? - ??????}
\end{abstract}

\pacs{?????? - ??????}
\maketitle



\textit{Introduction.} It is expected that a distant observer should see a black hole (BH) as a dark spot in the sky on the background of other bright sources, this dark spot is referred to as a 'BH shadow'.
Two projects are under way now to observe the shadow of supermassive BH which is supposed to be in the center of our galaxy: the
Event Horizon Telescope (http://eventhorizontelescope.org) and the BlackHoleCam (http://blackholecam.org). \textit{Ipso facto}, now great attention is focused on the study of various aspects of BH shadow, and investigations of shadow seems to become one of the most popular field of strong gravity scientific agenda. See selected classical and recent papers in \cite{Synge1966, Bardeen1973, Luminet1979, Chandra, Dymnikova1986, FalckeMeliaAgol2000, FrolovZelnikov2011, Perlick2014, Perlick2015, Interstellar2015, HiokiMaeda2009, Bambi2013, TsukamotoLiBambi2014, Zakharov2014, Cunha-PRL-2015, PerlickTsupkoBK2015, Rezzolla-Ahmedov-2015, JohannsenBroderick2016, Konoplya2016a, Ahmedov-add01, PerlickTsupko2017, Tsupko2017, Dokuchaev2017, Cite2018-Eiroa, Cite2018-Tsukamoto, Cite2018-Herdeiro, Johannsen-Wang-2016, Lu-Broderick-2014, Lu-Krichbaum-2018, Moscibrodzka-2009, Doeleman-2017, Mizuno-NatAstr-2018}.

As far as we live in the expanding universe, this expansion should influence on the size of a distant BH shadow. Nevertheless, a problem of calculation of the shadow in the expanding Friedmann universe is still not solved. To the best of our knowledge, all results are obtained only for the simplest case of Schwarzschild BH in de Sitter universe, described by the Kottler (sometimes referred as Schwarzschild--de-Sitter) metric \cite{Kottler-1918}. For a static form of the metric and for a static observer, the shadow size was calculated by Stuchl\'{i}k and Hled\'{i}k \cite{Stuchlik1999}. In subsequent papers \cite{Stuchlik2006, Stuchlik2007, Stuchlik2018} different moving observers are also considered. The first calculation of the shadow size as seen by comoving observer was performed by Perlick, Tsupko and Bisnovatyi-Kogan \cite{Perlick-Tsupko-BK-2018}. In particular, it is found that for a distant comoving observer the shadow shrinks to a finite value, but not to zero. For investigations of a particle motion and gravitational lensing in Kottler metric see, for example, \cite{Lake-Roeder-1977, Stuchlik1983, Islam1983, RindlerIshak2007, Hackmann-2008a, Hackmann-2008b, Lebedev2013, Lebedev2016}.

The difficulty of the analytical calculation of a shadow in a general case of an expanding Friedmann universe, described by Friedmann-Robertson-Walker (FRW) metric, contains two aspects. First, we need an adequate description of the BH embedded in the expanding universe. Second, we need to calculate light ray trajectories analytically in a strong gravity regime, and the crucial point for the analytical construction of the shadow is an existence of appropriate constants of motion. The most known ways to describe the BH embedded in the expanding universe are the Einstein-Straus model \cite{Einstein-Straus-1945, Einstein-Straus-1946, Schucking-1954} and McVittie metric \cite{McVittie1933, Nolan-1, Nolan-2, Nolan-3}, see also papers \cite{Gibbons-Maeda-2010, Hobson-2012a, Hobson-2012b} and review \cite{Carrera-Giulini-2010}. The Einstein-Straus model consists of Schwarzschild spacetime (called 'vacuole') matched to FRW metric with zero cosmological constant, see also \cite{Stuchlik1984, Balbinot-1988} for the case of a nonvanishing cosmological constant. Calculation of geodesics in this model is not simple: it can be done piecewise, with matching of Schwarzschild's part and FRW part at so called Sch\"{u}cking radius \cite{Schucking-1954, Schucker-2009, Schucker-2010}. In McVittie case, there is another difficulty: unlike the Schwarzschild--de-Sitter case, only angular momentum is conserved, and there is no second conserved quantity (energy). Therefore it is not possible to use only integrals of motion for calculation of the light trajectories, but we need to solve second-order differential equations \cite{Aghili-2017, Carrera-Giulini-2010}. Therefore, although there are investigations of gravitational lensing in McVittie metric in some approximations \cite{Aghili-2017, Lake-Abdelqader-2011, Piattella-PRD-2016, Piattella-Universe-2016, Faraoni-2017}, the possibility of analytical construction of shadow is, at least, questionable.

In this paper we propose an approximate method of calculation of the shadow in expanding Friedmann universe as seen by comoving observer which is appropriate for a general case. In realistic situations, we have the following two conditions: the observer is very far from the BH, at distances much larger than its horizon; the expansion is slow enough to be significant only on very large scales. Therefore we can neglect the influence of the expansion on a particle motion near the BH, and we can neglect the BH gravity during a long light travel to the observer if he is distant enough. Our idea is that in these approximations it becomes possible to reduce the problem of calculation of BH shadow to use of the angular diameter redshift relation \cite{Hobson, Mukhanov-book, Zeldovich-Novikov-book-2} with some effective 'physical size' of the shadow, which will give the correct approximate solution for general case. We have found this effective proper size of the shadow, and found the conditions under which this formula can be used. Actually, these conditions are satisfied for all shadows which can be considered for observation, excluding nearby (where expansion is negligible and therefore usual Schwarzschild formula can be applied). Our result is the dependence of the angular size of the shadow as a function of a BH redshift, for a given universe. This dependence is the approximate solution for general case and makes it easy to calculate the shadow in any multicomponent universe (with matter, radiation and dark energy).

Applying this formula to the real universe with known parameters, we find that presence of matter leads to unbound increase of the shadow radius with increase of the BH redshift. For large enough supermassive BH at cosmological distances, it is revealed that the shadow size can approach to the size of the BH shadow in the center of our galaxy.\\


\textit{Description of method.} Let us consider the Schwarzschild BH embedded in an expanding universe filled by matter, radiation and dark energy. Expansion of the universe is described by the Hubble parameter $H(t)$, and we denote the present day value as $H_0$. The mass of a BH is $M$, the Schwarzschild radius is $R_S=2m$, mass parameter is $m=MG/c^2$. An observer comoving with the cosmic expansion is situated far from the BH, his radial coordinate is $r_O \gg m$ (BH is in the origin of coordinates).

Let us suppose that we can introduce a radial coordinate $r_1$ such that the following relation holds:
\begin{equation}
m \ll r_1 \ll r_O \, .
\label{eq:regions}
\end{equation}
We assume that coordinate $r_1$ is large enough in comparison with the BH horizon to neglect the BH gravity. At the same time $r_1$ is small enough in comparison with observer's coordinate to neglect the cosmic expansion. Condition (\ref{eq:regions}) allows us to identify the two regions of space:

(i) the region near the BH, $r < r_1$, and

(ii) the region far from the BH, $r > r_1$.

In the first region we can neglect the effects of expansion in comparison with BH gravity. In the second region the cosmic expansion predominates, and we can neglect BH gravity.

Angular size redshift relation can be applied for the second region. A crucial thing we need to know for that is a physical linear size of the object. Let us find 'linear' size of the shadow, if we assume that it is observed from the distance $r_1$. According
the Synge's \cite{Synge1966} formula for the angular radius $\alpha_{\mathrm{sh}}$ of the shadow
of the Schwarzschild BH we can write:
\begin{equation}
\sin ^2 \alpha_{\mathrm{sh}} \, = \,
\frac{27m^2 (1-2m/r_1)}{r_1^2} \, \simeq \, \frac{27m^2}{r_1^2} .
\end{equation}
Since it is assumed that the cosmic expansion is small enough, we can consider that at radius $r_1$ the space-time is almost flat, and write that linear radius $R_{\mathrm{sh}}$ of the shadow equals to:
\begin{equation}
R_{\mathrm{sh}} \, \simeq \, r_1 \, \sin \alpha_{\mathrm{sh}} \, = \, 3 \sqrt{3} m \, .
\end{equation}
Obviously, this value is simply equal to the critical impact parameter corresponding to photons approaching the photon sphere $r_{\mathrm{ph}} = 3m$. Note that near the BH, at radius $r_1$, we don't differ between static and comoving observers because the expansion near the BH is small.

Now let us consider the second region in the following way: distant observer with position $r_O$ observes the image of the object which is located at $r_1$ and has the linear radius $R_{\mathrm{sh}}$. Due to condition (\ref{eq:regions}) we will neglect $r_1$ in comparison with $r_O$ and consider that $r_O-r_1 \simeq r_O$, so for distant observer the object of the linear radius $R_{\mathrm{sh}}$ is located at the same place as the BH. Then, to use angular size redshift relation we shall consider that the observer sees the shadow of BH which is situated far away with redshift $z$. The value of this redshift is estimated by observations of the spectra of a galaxy or quasar surrounding a supermassive BH.

Angular size redshift relation is presented in literature very well, see, for example, \cite{Hobson, Mukhanov-book, Zeldovich-Novikov-book-2}, see also classical papers \cite{Mattig-1958, Zeldovich-1964, Dashevsk-Zeldovich-1965}. According to definition of angular diameter distance $D_A$, we write:
\begin{equation}
D_A = \frac{L}{\Delta \theta} \, ,
\end{equation}
where $L$ is the proper diameter of the object, and $\Delta \theta$ is the observed angular diameter. For a given universe, angular diameter distance is a known function of the redshift $z$ (we restrict ourselves by the flat case universe):
\begin{equation}
D_A(z) = \frac{c}{(1+z) H_0} Int(z) \, ,
\end{equation}
where
\begin{equation}
Int(z) = \int \limits_0^z \left( \Omega_{m0} (1+\tilde{z})^3 + \Omega_{r0} (1+\tilde{z})^4 + \Omega_{\Lambda 0} \right)^{-1/2} d\tilde{z} \, ,
\end{equation}
$H_0$ is the present day value of the Hubble parameter $H(t)$, and $\Omega_{m0}$, $\Omega_{r0}$, $\Omega_{\Lambda 0}$ are the present day values of density parameters for matter, radiation and dark energy correspondingly.

We find that the angular diameter is the following function of redshift:
\begin{equation}
\Delta \theta (z) = L \, \frac{H_0}{c} \, \frac{1+z}{Int(z)} \, .
\end{equation}

In this paper we are working with the angular radius, therefore we will use this formula with the proper radius and the angular radius:
\begin{equation}
\alpha_{\mathrm{sh}}(z) = R_{\mathrm{sh}}  \, \frac{H_0}{c} \, \frac{1+z}{Int(z)} \, .
\end{equation}
Substituting $R_{\mathrm{sh}} = 3 \sqrt{3} m$, we obtain the final formula for the angular radius of the shadow as:
\begin{equation} \label{main-result}
\alpha_{\mathrm{sh}}(z) = 3\sqrt{3}m \, \frac{H_0}{c} \, \frac{1+z}{Int(z)} \, .
\end{equation}
This formula allows to calculate the size of the shadow as a function of the BH redshift in universe with given parameters $H_0$, $\Omega_{m0}$, $\Omega_{r0}$, $\Omega_{\Lambda 0}$.

A limiting case of small $z$ can be treated simply. We obtain that $Int(z) \simeq z$ (remind that $\Omega_{m0} + \Omega_{r0} + \Omega_{\Lambda 0} = 1$), and the angular radius equals to
\begin{equation} \label{small-z}
\alpha_{\mathrm{sh}}(z) = 3\sqrt{3}m \, \frac{H_0}{c} \, \frac{1}{z} \, .
\end{equation}
Taking the distance to the BH as $D_d \simeq cz/H_0$, which is correct for $z \ll 1$, we recover the formula for the shadow size at large distances in Schwarzschild metric in the universe, when the expansion may be neglected, as
\begin{equation} \label{small-z-2}
\alpha_{\mathrm{sh}}(z) = \frac{3\sqrt{3}m}{D_d} \, .
\end{equation}
For example, let us consider the supermassive BH in M87. Taking $z=0.004283$ \cite{M87-redshift} and $M = 6.2 \times 10^9 M_\odot$ \cite{Perlick2015}, we obtain the angular diameter about 35 $\mu$as. Note that our approximation does not allow to use very small values of $z$, because the angle of the shadow should still be small, and the condition (\ref{eq:regions}) should hold. But for all realistic situations these conditions are valid.

Now let us consider some particular cases.



\textit{Black hole in de Sitter universe.} First of all, we investigate the shadow of the BH in de Sitter universe as seen by comoving observer. For de Sitter universe we have $\Omega_{\Lambda 0} = 1$, $H(t)=H_0=\text{const}$ and $Int(z) = z$. We obtain:
\begin{equation} \label{sds0}
\alpha _{\mathrm{sh}} =
3\sqrt{3}m \, \dfrac{H_0}{c} \, \frac{1+z}{z} \, .
\end{equation}

Calculation of the shadow for this case was performed exactly in \cite{Perlick-Tsupko-BK-2018} where we considered the Kottler metric for the description of the BH in de Sitter universe. Let us demonstrate that approximate calculations of the present paper coincide with exact results of paper \cite{Perlick-Tsupko-BK-2018} simplified according to used approximations.

Formula (51) from \cite{Perlick-Tsupko-BK-2018} written for distant comoving observer ($r_O \gg m$) is the following:
\begin{equation}
\alpha _{\mathrm{sh}} \approx
\dfrac{3 \sqrt{3} \, m}{r_O}  \left( \sqrt{ 1 - \dfrac{27H_0^2m^2}{c^2}} + \dfrac{H_0 r_O}{c}  \right)
\, .
\end{equation}
In the present paper we assume that the expansion is negligible near the BH, it means that $H_0m/c \ll 1$ and the second term inside the square root can be neglected:
\begin{equation}  \label{eq:sds2}
\alpha _{\mathrm{sh}} \approx
\dfrac{3 \sqrt{3} \, m}{r_O}  \left( 1 + \dfrac{H_0 r_O}{c}  \right)
\, .
\end{equation}
Here $r_O$ is coordinate of comoving observer in static representation of the Kottler metric. To use comoving coordinates $\tilde{r}$ and $\tilde{t}$, for large distances from BH, we should substitute
\begin{equation}  \label{eq:sds3}
r_O = \tilde{r}_O e^{H_0 \tilde{t}_O}
\end{equation}
(see \cite{Perlick-Tsupko-BK-2018} for details). Connection between comoving coordinate $\tilde{r}$ and redshift for constant $H(z)=H_0$ is \cite{Hobson}
\begin{equation} \label{eq:sds4}
\tilde{r} = \frac{c}{e^{H_0 \tilde{t}_0}} \int \limits_0^z \frac{dz}{H(z)} = \frac{c}{e^{H_0 \tilde{t}_0}} \frac{z}{H_0} .
\end{equation}
Here $\exp(H_0 \tilde{t}_0)$ is the present day value of the scale factor, and here the time $\tilde{t}_0$ corresponds to the time of observation $\tilde{t}_O$ in eq. (\ref{eq:sds3}). Using (\ref{eq:sds3}) and (\ref{eq:sds4}) in (\ref{eq:sds2}), we recover (\ref{sds0}).

Taking $z \to \infty$ in (\ref{sds0}), we obtain
\begin{equation} \label{sds}
\alpha _{\mathrm{sh}} \to
3\sqrt{3}m \, \dfrac{H_0}{c} \, .
\end{equation}
This limit was found in the previous paper \cite{Perlick-Tsupko-BK-2018}, see formula (40).

The presence of expansion due to cosmological constant leads to phenomenon that for a distant comoving observer the shadow shrinks to finite value (\ref{sds}), but not to zero. Below we show that the presence of the matter leads to increase of the shadow for large $z$.\\

\begin{figure}
	\begin{center}
		\includegraphics[width=0.46\textwidth]{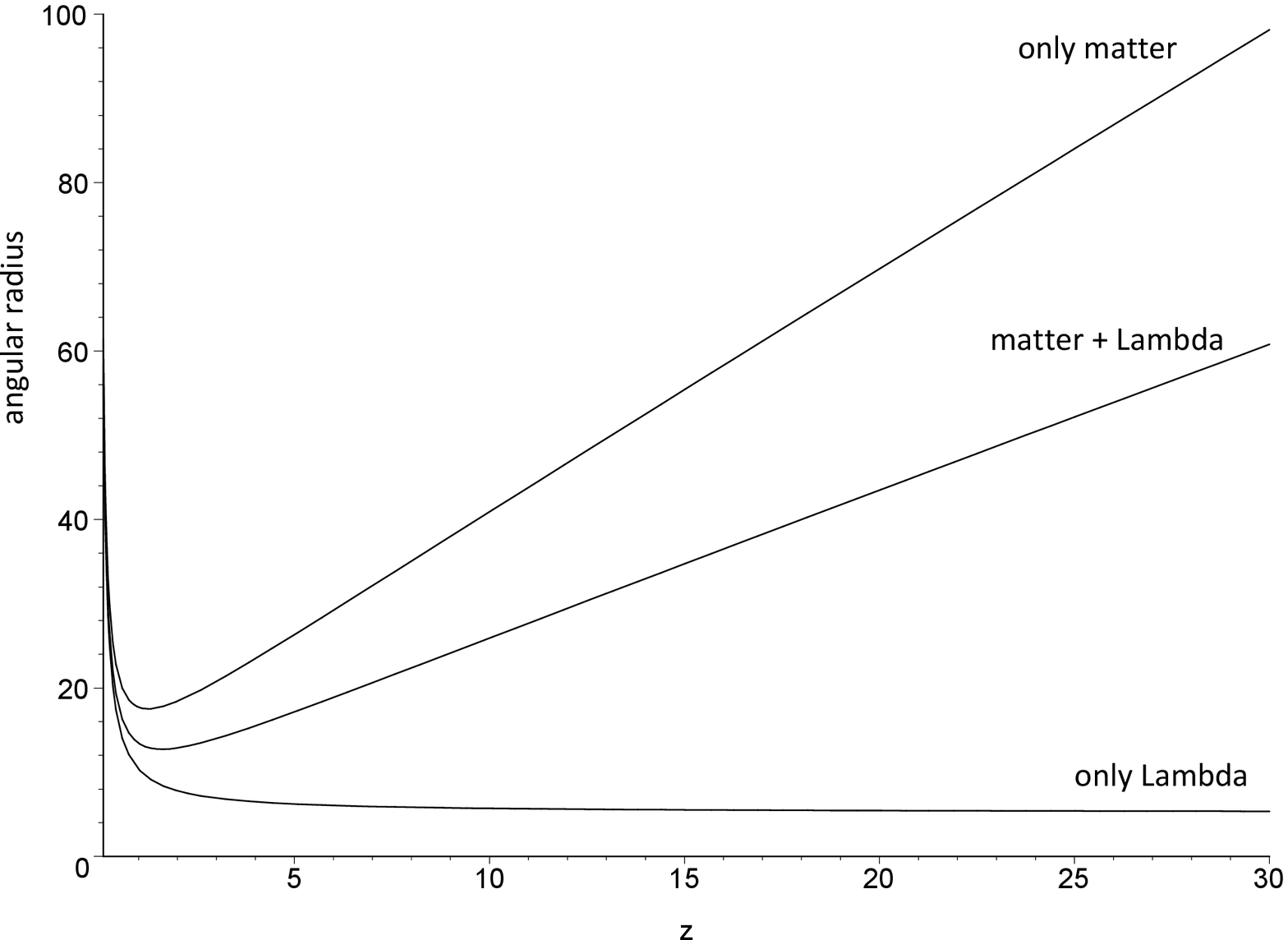}
	\end{center}
	\caption{Dependence of angular radius $\alpha_{\mathrm{sh}}$ on $z$ for different spacetimes: de Sitter model without matter ($\Omega_{m0}=0$, $\Omega_{\Lambda 0}=1$), Einstein--de-Sitter spacetime filled by matter only ($\Omega_{m0}=1$, $\Omega_{\Lambda 0}=0$) and mixed case with matter and dark energy presence ($\Omega_{m0}=0.3$, $\Omega_{\Lambda 0}=0.7$). Angular radius is presented in units of $mH_0/c$.} \label{fig:curves-z-30}
\end{figure}

\begin{figure}
	\begin{center}
		\includegraphics[width=0.46\textwidth]{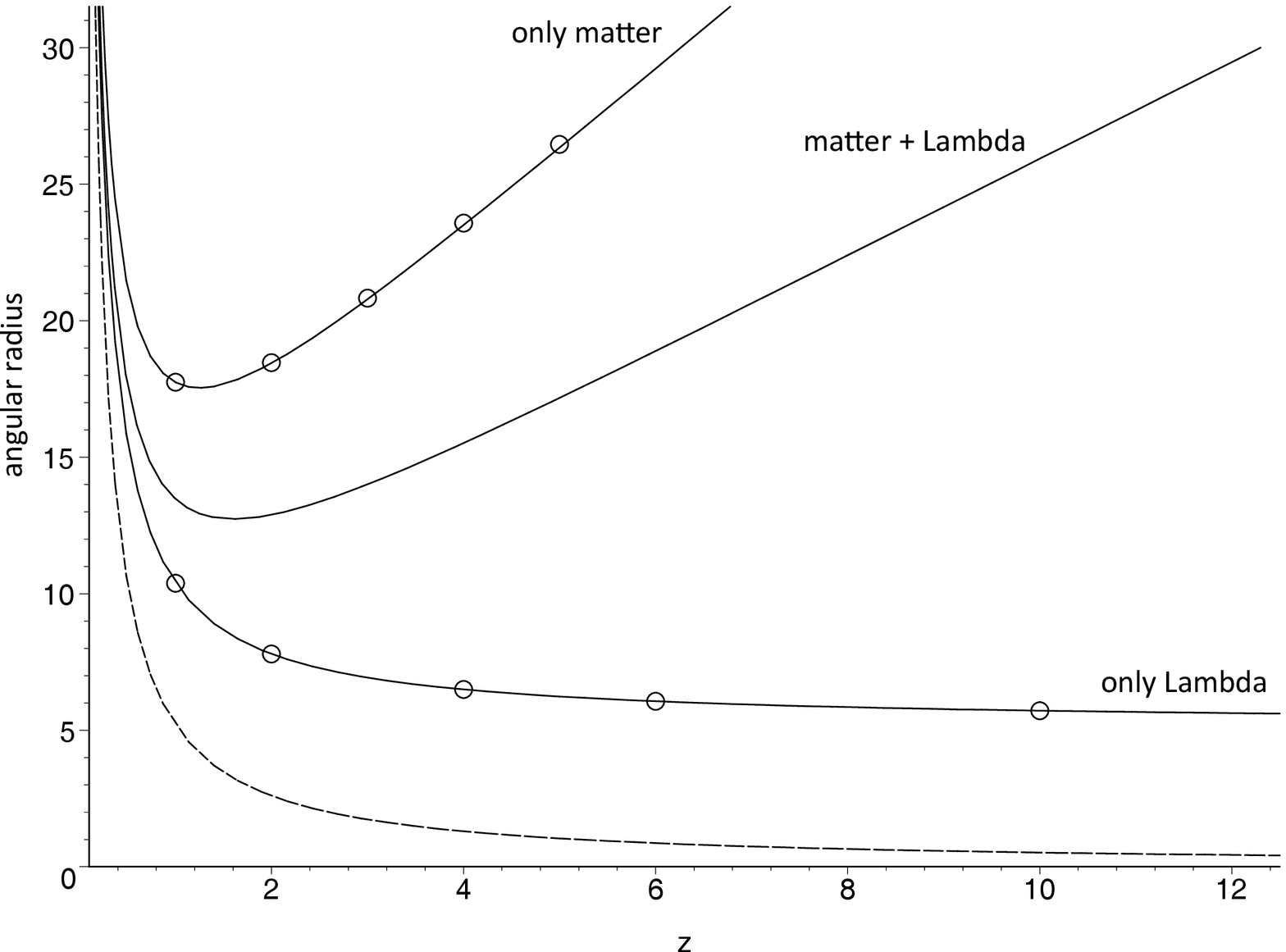}
	\end{center}
	\caption{Zoom of previous figure. Here we have also shown the additional dashed curve plotted according to formula (\ref{small-z}). Formula (\ref{small-z})
is only valid for small $z$, but in the whole range of $z$ it can be interpreted as the
Schwarzschild case without cosmological expansion.\footnote{We are thankful to anonymous referee for this remark.} It can be understood from transformation of (\ref{small-z}) to (\ref{small-z-2}). Circles on curves show the results of numerical calculation of the shadow size by the integration of the null geodesics in McVittie spacetime, see details in the text. } \label{fig:curves-z-12}
\end{figure}

\textit{Einstein--de-Sitter spacetime.} Now let us consider matter dominated universe, namely a flat universe filled by dust \cite{Hobson, Mukhanov-book}. Einstein--de-Sitter spacetime has the following parameters: $\Omega_{m0}=1$, $\Omega_{r0}=0$ and $\Omega_{\Lambda 0}=0$. We obtain:
\begin{equation}
\alpha_{\mathrm{sh}}(z) = \frac{3\sqrt{3}m}{2} \, \frac{H_0}{c} \, \frac{1+z}{[1 - (1+z)^{-1/2}]} \, .
\end{equation}
Minimum of $\alpha_{\mathrm{sh}}(z)$ corresponds to $z=5/4$. We reveal that angular size increases unboundedly with increase of the BH redshift (similar to behaviour of angular size of another object in such universe).\\

\textit{Real multicomponent universe.} The main interest is in results for universe with parameters known from modern observations. To model the real universe, we use the following parameters: $\Omega_{m0}=0.3$, $\Omega_{r0}=0$ and $\Omega_{\Lambda 0}=0.7$. Picture of these particular cases is presented in Figs. \ref{fig:curves-z-30}-\ref{fig:curves-z-12}.

For $z=10$ we find that $\alpha_{\mathrm{sh}}  \simeq 26 \, mH_0/c$. With $M=10^{10} M_\odot$ (compare with M87) and $H_0 = 70$ km/(sec$\cdot$Mpc) we obtain that $\alpha_{\mathrm{sh}} \simeq 0.6 $ microarcseconds. With larger mass, $M=10^{11} M_\odot$, we have $\alpha_{\mathrm{sh}} \simeq 6 $ microarcseconds. We conclude that angular diameter of BH at cosmological distances may reach 12 $\mu$as. For comparison, the shadow size of BH Sgr A* ($M = 4.3 \times 10^6 M_\odot$) can be estimated as 53 $\mu$as \cite{Perlick2015}. Observations of very massive black holes at high redshifts can be found, e.g., in \cite{Quasar-1, Quasar-2}.

\textit{Comparison with exact numerical solution in McVittie metric.} To further verify our result, we have obtained the shadow size by numerical integration of the null geodesic equations in McVittie metric written in 'comoving' coordinates as ($G=c=1$):
\begin{equation} \label{eq:McVittie}
ds^2 = - \left( \frac{1-\mu}{1+\mu} \right)^2 dt^2 + (1+\mu)^2 a^2(t) (dr^2 + r^2 d\Omega^2) , 
\end{equation}
\begin{equation}
\text{where} \; \mu = \frac{m}{2a(t)r}, \; d \Omega ^2 = \mathrm{sin} ^2 \vartheta \, d \varphi ^2 + d \vartheta ^2 \, ,
\end{equation}
$a(t)$ is the scale factor. Equations of motion located in literature and usually used for numerical integration are derived with using another coordinates (sometimes called 'static'), see \cite{Nolan-3, Aghili-2017, Hobson-2012a, Hobson-2012b, Lake-Abdelqader-2011}. For our purpose, we have to find the equations of motion in 'comoving' coordinates. For simplicity, in the metric (\ref{eq:McVittie}) and in the rest of the article, we write these coordinates without a tilde.

Using Lagrange function $\mathscr{L} = g_{\mu\nu} \dot{x}^\mu \dot{x}^\nu$ ('dot' means the differentiation with respect to an affine parameter $\lambda$), we have derived the equations of motion for coordinates $t$, $r$, $\varphi$, for motion in equatorial plane of metric with constant angular momentum $L$. Since all non-zero metric coefficients of (\ref{eq:McVittie}) depends on both $t$ and $r$, the resulting equations are rather cumbersome to be presented here.

In numerical integration, we send the light rays to the past from chosen observer's position, and seek for the boundary between absorbed and flown by rays, which gives us the numerical size of the shadow. We choose the explicit form of $a(t)$, the Hubble value $H_0$ and observer's position $r(\lambda_0)$; for simplicity $t(\lambda_0)$ is chosen to give $a(t_0)=1$; we also fix initial negative value $\dot{t}(\lambda_0)$. To vary the angle of emission of light ray we vary $\dot{r}(\lambda_0)$. Transformation between pair of comoving coordinates $r(\lambda_0)$, $t(\lambda_0)$ and $z$ is performed by formula similar to (\ref{eq:sds4}), with corresponding change of notation and expressions for $a(t_0)$ and $H(z)$.


We have made the calculations for the cases of cosmological term only and matter only, where the scale factor has simple dependences on time. To satisfy our approximations, we have used small values of $H_0$ (see below) and large values of $r(\lambda_0) \gg m$ ($G=c=1$, and all variables below are non-dimensional, with $m=1$, indicated as 'nd'):

(a) de Sitter metric, $a(t)=\exp(H_0^{nd} t^{nd})$, $H_0^{nd}=0.001$, $t^{nd}(\lambda_0)=0$, different $r^{nd}(\lambda_0) \simeq 10^3 - 10^4$ (corresponds to different $z$);

(b) Einstein-de Sitter, $a(t) = (3 H_0^{nd} t^{nd}/2)^{2/3}$, $H_0^{nd}=0.0001$, $t^{nd}(\lambda_0)=2/3H_0^{nd}$, $r^{nd}(\lambda_0) \simeq 10^3 -  10^4$. Note that in this section $r(\lambda)$ is in the units of $GM/c^2$.

Points found in result of numerical integration are shown in Fig.\ref{fig:curves-z-12}. We conclude that our approximate results agree with the numerical calculation of the shadow size.

Note that for real universe with $H_0 = 70$ km/(sec$\cdot$Mpc) the non-dimensional Hubble parameter $H_0^{nd}$ is even smaller: $H_0^{nd} = H_0GM/c^3$, which gives $10^{-13}$ for $M=10^{10} M_\odot$, and $10^{-17}$ for $M=10^{6} M_\odot$. Therefore our approximations will certainly be satisfied.


\textit{Concluding remarks}

(i) As discussed above, it seems to be very difficult to calculate exactly the shadow size in a general case of the expanding universe. Here we have derived the approximate formula for the shadow size as seen by a distant comoving observer using angular diameter redshift relation and effective linear size of the shadow, see (\ref{main-result}). Formula can be easily applied for calculation of the shadow in any multicomponent universe (with matter, radiation and dark energy), in approximation that comoving observer is far from a BH. In particular case of Schwarzschild--de-Sitter (when the expansion is driven by cosmological constant only) our approximate results agree with known exact solution. Our formula also gives the correct results for a case of small $z$. Our results also agree with the shadow size obtained by numerical integration of light rays in McVittie metric. Such way of calculation of the shadow is used for the first time.

(ii) Spectacular thing following from our consideration is
that in presence of the matter component the shadow size of supermassive BHs at cosmological distances may reach the values which are only one
order of magnitude less than the shadow size in the center of our galaxy and the present sensitivity limits. Moreover, BHs at larger values of $z$ will lead to even larger size of the shadow. Recent radio observations of absorption background 21 cm line have revealed some properties of universe up to $z \simeq 20$ \cite{21cm}. Outstanding future project 'James Webb Space Telescope' (https://jwst.nasa.gov) designed for observations at large $z$ in infrared would be able to discover the objects presumably of such large redshifts. Subsequent observations of these objects with high resolution can reveal the shadow of supermassive BH at large cosmological distances.

(iii) We have to stress that the proposed possibility of observation of such shadow contains in itself all the problems that there are in the observations of the shadow in the center of our galaxy. For the observability of the shadow it is necessary not only that its angular radius is big enough but also that there is a backdrop of  sufficiently bright light sources against which the shadow can be seen as a dark spot. For example, emission of matter surrounding the BH would partially obscure any shadow. Shadow in the center of our galaxy cannot be seen in optical band, and observations are supposed to be in submillimeter wavelength. Speaking about the observation of the shadow of extragalactic BH in infrared band, it should be noted that this infrared radiation is the product of optical emission redshifted by the cosmological expansion. Therefore we need to consider the BH which is similar to M87 where the radiation in optical band is not absorbed. In the case of accretion disc presence around the supermassive BH, we need to consider a situation when the observer isn't located in the equatorial plane of the BH. In this work we consider only non-spinning BHs. Although BHs are likely expected to be rotating, the main size of the shadow for a rotating BH is still set largely by the mass.

Comparing current observations of BH shadow in galactic center and possible future observations of shadow of cosmological BHs, we would like to mention the following positive feature of the last ones. In case of observation of supermassive BH in galactic center the optical observations are not possible because optical radiation is screened by dust. Therefore observations are performed in sub-mm range. In case of quasars, optical observations are possible and performed. Since there is three order of magnitude difference in wavelengths between optical and mm band, using of VLBI technology could provide three order of magnitude better resolution. This will be definitely enough to observe the shadow of cosmological BHs.

\textit{Acknowledgments}

We are thankful to Volker Perlick for many useful discussions. This work is financially supported by Russian Science Foundation, Grant No. 18-12-00378.

\bibliographystyle{ieeetr}

\end{document}